\documentstyle[multicol,graphics,aps]{revtex}
\begin{document}
\title{Time-dependent Effects in the Metallic 
Phase in Si-MOS: Evidence for Non-Diffusive Transport}
\author{V.\ M.\ Pudalov$^a$, G.\ Brunthaler$^b$, A.\ Prinz$^b$,  
G.\ Bauer$^b$, B.\ I.\ Fouks$^c$}
\address{$^a$P.\ N.\ Lebedev Physics Institute, Leninsky prosp. 53, Moscow 119324, Russia}
\address{$^b$Institut f\"{u}r Halbleiterphysik, Johannes Kepler Universit\"{a}t,
Linz, A-4040, Austria}
\address{$^c$Institute for Radioengeneering and Electronics, Mochovaya str. 18,
Moscow 103907, Russia}
\date{\today}
\maketitle

\begin{abstract}
We have found that the conduction in Si-MOS structures has a substantial
imaginary component in the metallic phase for the density range
$ 6\times n_c > n > n_c$, where $n_c$ is the critical density of the 
metal-insulator transition.
For high mobility samples, the corresponding delay (or advance) time, 
$\tau  \sim (0.1 - 10)$\,ms, 
increases exponentially as density 
and temperature decrease. In very low mobility samples, at temperature of  0.3K,
the time-lag in establishing  the equilibrium resistance reaches 
hundreds of seconds.
The delay (advance) times are approximately $10^2-10^8$ times larger than 
the overall $RC$-time of the gated structure.
These results give evidence for a non-Boltzmann character
of the transport in the low-density metallic phase. 
We relate the time-dependent effects to tunneling of carries 
between the 2D bulk and localized states. 
\end{abstract}

\begin{multicols}{2}
The metal-insulator transition observed in different two dimensional carrier 
systems \cite{experim} is currently in the focus of interest. 
A number of models put forward for its explanation 
\cite{interaction,spingap,maslov99,dassarma}
span from a non-Fermi-liquid state
\cite{interaction}
to interface traps physics \cite{maslov99,dassarma}. 
In the latter models, the strong exponential 
drop in the resistivity at low temperature
is a transient temperature effect only.
On the experimental side, from measurements at high carrier density
\cite{Gmax},
the exponential drop was proven not to be related to 
the ground state conduction, at least 
for high carrier densities $n \geq (10-15) \times n_c$. 

The involvement of interface traps into the transport may be revealed 
by studying charging effects, time lag,  noise character etc.
We present here data evidencing that the time-dependent 
effects are  essential in the metallic phase, 
even for densities  6 times the critical  one, $n_c$. 
We studied in detail four samples with different peak mobility, 
Si-11 ($\mu^{peak}=39,000$\,cm$^2$/Vs), Si-22 ($\mu^{peak}=33,000$\,cm$^2$/Vs), 
Si-4/32 ($\mu = 8400$\,cm$^2$/Vs), and Si-52 ($\mu = 1300$\,cm$^2$/Vs). 
For the first three samples, an Al gate film was deposited onto the SiO$_2$ 
layer followed by a post-metallization anneal.  
The 
density of trapped carriers (estimated from the threshold voltage 
\cite{JETPL99a}), 
was $(2-3)\times 10^{10}$\,cm$^{-2}$ for the first two samples, and 
$21\times 10^{10}$\,cm$^{-2}$  for the third one. 
In the lowest mobility sample, we increased intentionally the 
amount of disorder by thermal evaporating  an Inconel gate 
without a subsequent anneal. Beyond the substantial 
decrease in the mobility, the amount of trapped carriers increased up 
to $\sim 60\times 10^{10}$\,cm$^{-2}$.
All samples  were of the same Hall-bar geometry,
$5 \times 0.8$\,mm$^2$, \cite{JETPL99a} with gate oxide 
thickness of $d_{\rm SiO_{2}} = 200 \pm 20$\,nm, 
aspect ratio of $w/l=0.32$, and
corresponding capacitance between the gate 
and 2D layer $C \approx 690$\,pF.
The potential and current contacts to the 2D channel
were lithographically defined and made by thermal diffusion of phosphorus.
The overall  device $RC$-time, including  contact resistance
was of the order of $(1 - 10)\,\mu$s, 
and was expected to contribute a negligibly small imaginary 
component to the sample ac-conductance. 

Four-terminal ac-transport measurements were carried out 
in the frequency range  0.3 to 30\,Hz with a quadrature 
lock-in amplifier. In order to eliminate the influence of the resistance 
of potential probes, 
we used a battery operated electrometric 
preamplifier with an input current less than 1\,pA. The amplifier 
phase-frequency characteristic was verified not to contribute 
to the studied effects.
In all samples, we found the time-dependent 
effects to persist 
in the metallic phase, far above the critical density. 
In high mobility samples, 
the characteristic times were in the ms-range and 
were measured from a
phase shift $\varphi$ between the voltage drop, $V_x$, 
and the source-drain current, $I_x$, as well as
from the  frequency dependence $\varphi(F)$.
In low mobility samples, the characteristic times were of order of 
1-100\,s and were measured directly, by applying a small voltage 
step on the top of the constant gate voltage and 
measuring the transient voltage $V_x$ at a constant current $I_x$.
In the following, we define the ``resistivity'', $\rho$ as
the in-phase component, $Re((V_x/I_x) \times w/l )$.

{\bf Measurements with high mobility samples.} Figure~1 shows 
the phase shift between the ac voltage $V_x$
and the current $I_x$, 
measured in a high mobility sample
at a frequency of 3.8\,Hz,
as a function of carrier density. 
The phase shift emerges as the density  decreases below 
$6\times10^{11}$\,cm$^{-2}$.
This is about 6 times  larger than the critical density, 
$n_c=0.95 \times 10^{11}$\,cm$^{-2}$ for the metal-insulator 
transition in this sample.

The lower inset of Fig.~1a shows that, as  density decreases, 
the phase shift first is negative  
(which corresponds to the voltage delay), then
becomes positive (voltage advancing), and, finally, becomes again 
negative close to the critical density. 
The phase shift increases linearly with ac-current frequency $F$
and may thus be interpreted as a time
delay (or advance, correspondingly), $\tau = \varphi /(2\pi F)$. 
The upper inset in Fig.~1\,a shows  delay time (= $|\tau |$) calculated from the slope 
of the frequency characteristics, at a fixed temperature of 290\,mK, and
over the range of high densities where $\phi < 0$.  
As temperature decreases, the phase shift displays more and more 
pronounced oscillations as a function of density. The phase shift 
is reproducible during the same cool-down, however, in different 
cool-downs the oscillatory details varied 
on the density scale. 
\begin{figure}
\begin{center}
\resizebox{7.8cm}{10cm}{\includegraphics{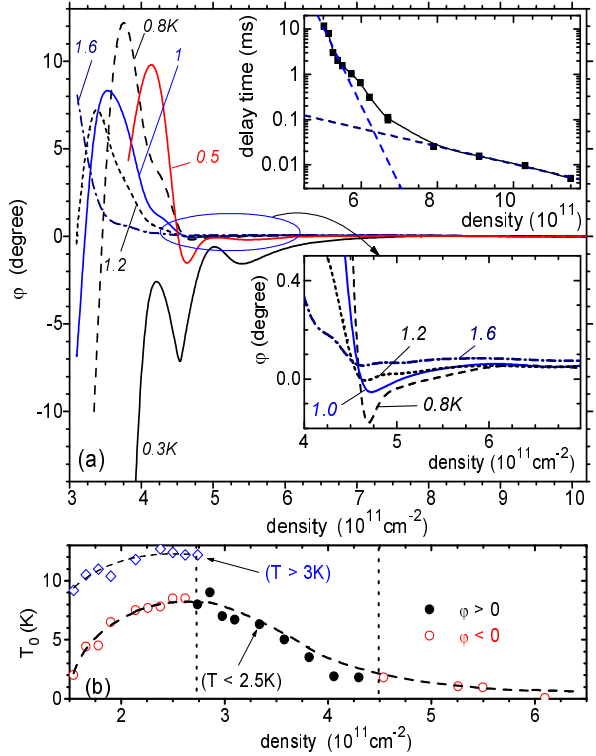}}
\begin{minipage}{7.5cm}
\vspace{0.2cm}
\caption{(a) Phase shift between ac-voltage $V_x$ and current $I_x$ 
measured with sample Si-22 at frequency 3.8\,Hz for 
6 temperatures (indicated on the figure). 
The range of  $n = (4-7)\times 10^{11}$\,cm$^{-2}$ is blown up 
on the lower inset. Upper inset displays 
delay time vs density, measured from the 
frequency dependence of the phase shift at $T=0.29$\,K. 
(b) ``Activation energy'', $T_0$, for low and high temperature
ranges. Vertical dashed lines $n= n_i$ 
separate the regions of positive and negative $\varphi $.}
\label{fig1}
\end{minipage}
\end{center}
\end{figure}
\vspace{-0.2cm}
Figures~2\,a and 2b show the phase shift and the resistivity 
for sample Si-22
as a function of temperature for 
eight fixed densities.
It is remarkable that
{\em the strong exponential drop in resistivity develops in the same 
ranges of densities and temperatures as the phase shift does}, 
although we can not simply relate the two effects to each other. 

As follows from Figs.~1a and 2a,
both, $\tau $ and $\varphi$ decay  
about exponentially
with density and temperature, 
\begin{equation}
\tau \propto f_1(n,T) \exp (T_0(n)/T).
\end{equation}
The prefactor $f_1$ oscillates as a function of density 
(and of temperature) changing sign at "node" values, $n_i$. 
The definition of the slope, $T_0$,
is illustrated by the dashed tangent lines 
in the lower inset of Fig.~2a.

$T_0(n)$ is not constant over entire temperature range:
it is large for high temperatures, and decreases for lower
\begin{figure}
\begin{center}
\resizebox{8.0cm}{10cm}{\includegraphics{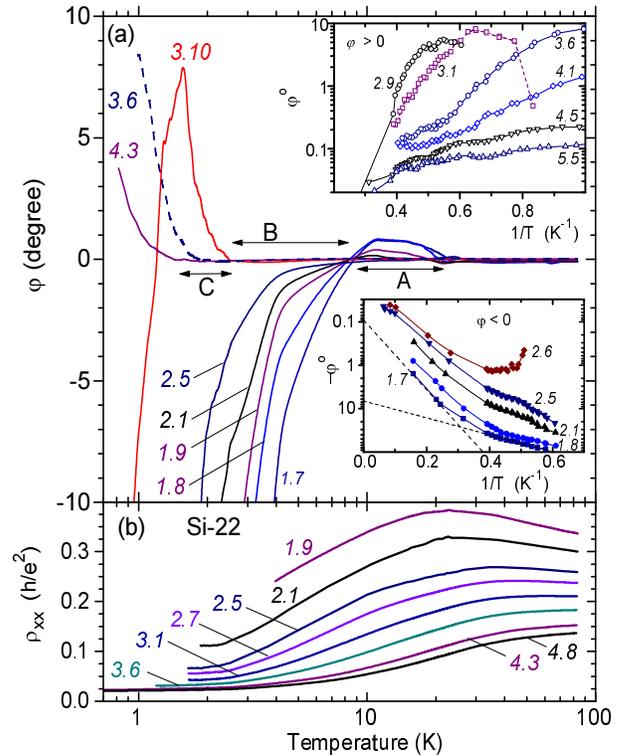}}
\begin{minipage}{7.5cm}
\vspace{0.2cm}
\caption{(a) Temperature dependence of  the phase shift,
$\varphi $,
measured at 3.8\,Hz for 
eight different densities (indicated on the figure in units 
of $10^{11}$\,cm$^{-2}$). Lower and upper insets 
are the ``Arrhenius plots''
for $\varphi$  vs $1/T$: 
lower  inset  is for the negative $\varphi$ over the temperature 
range ``B''; upper inset is for the positive 
$\varphi$ over the  range ``C''.
(b) Temperature dependence of the 
resistivity for the same eight densities
as in (a).}
\label{fig2}
\end{minipage}
\end{center}
\end{figure}
\vspace{-0.2cm}
 $T$'s.
In the vicinity of the nodes $n_i$ and for low temperatures,
the slope tends to vanish 
which is simply affected by the oscillatory 
behavior of $f_1(T)$. Therefore, the narrow density ranges 
around the nodes were ignored in the calculations
of $T_0$. 
The resulting density dependence 
$T_0(n)$ is shown in Fig.~1b, evaluated separately
for high ($T > 2.5$\,K) and low ($T < 2.5$\,K) temperatures.
Despite Eq.~(1) describes the data 
very roughly, an important conclusion can be drawn immediately:
$T_0$ does not decrease to 0 for the nodes $n=n_i$ 
and develops smoothly from the ranges of $\varphi >0$ 
to those of $\varphi <0$.
This means that the nodes are related to the prefactor 
$f_1(n)$ rather than to the exponential factor.
For high densities, $T_0$ decays 
steeper than $\propto (n - n_0)^{-1}$, 
as shown in Fig. 1\,b, thus causing
an exponential decay of $\tau $ for high densities.

{\bf Low mobility sample}. 
In the low mobility sample Si-52, the time-dependent 
effects are much stronger
and manifest themselves in a transient voltage
between potential probes when the gate voltage $V_g$ changes by a small step
$\Delta V_g \ll V_g$. Typical transient 
curves  {\it 1 - 5} of $\Delta \rho (t) = \rho(t)-\rho(0)$ 
normalized by $\Delta \rho_0 = \rho(\infty ) - \rho(0)$ 
are shown in Fig.~3a, for 5 different densities. 
The curves were fitted with exponential functions
$\rho_0 \exp(-t/\tau_d)$ (shown by dashed curves), 
from which the time lag $ \tau_d$ was obtained. 
At some, rather arbitrary densities, the transient curves 
were non-monotonic with oscillations (curve {\it 6}), 
or jumps (curves {\it 7, 8, 9}).

\begin{figure}
\begin{center}
\resizebox{6.5cm}{6cm}{\includegraphics{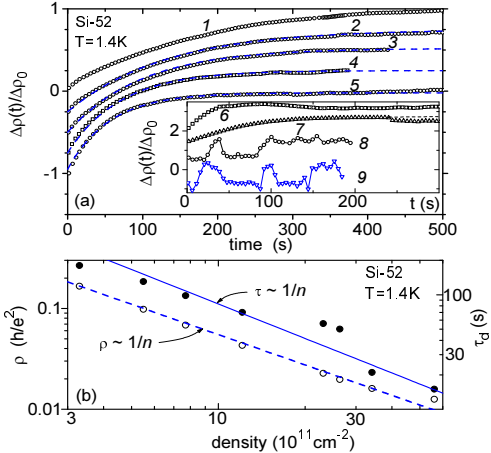}}
\begin{minipage}{7.5cm}
\vspace{0.2cm}
\caption{(a) $\Delta \rho (t)$, normalized by $\rho (\infty ) -\rho(0)$. Curves {\it 1-5} 
are shifted vertically for clarity. The curves {\it 1-5} correspond to the 
carrier densities 
3.3, 5.5, 7.7, 12.1, 23.1 (in units of $10^{11}$\,cm$^{-2}$). 
Curves {\it 6-9} ($n= 34, 26, 23$) 
in the inset show examples of non-monotonic 
behavior.   
(b) Time lag $\tau_d $ (right $y$-axis) and 
$\rho $ (left $y$-axis) vs carrier density.}
\label{fig3}
\end{minipage}
\end{center}
\end{figure}
\vspace{-0.2cm}
The time lag $\tau_d$ and resistivity $\rho $ 
are plotted in Fig.~3b  vs carrier density.
Their ratio,
$\tau_d /\rho $, is again $10^8$ times larger than the sample capacitance
pointing to its irrelevance.

{\bf Discussion}. 
The characteristic times, even for high mobility samples,
are of the order of $10^{-4}-10^{-2}$s  and, 
therefore, can not be associated 
with any  $RC$ time constants of the sample. 
Having the typical sample parameters 
$R \sim (10^3 - 10^4)$\,Ohm, and $C = 0.7\times 10^{-9}$\,F, one  can 
hardly find in the sample either a capacitance 
$\sim 10^{-4}$\,F, or a resistance  $\sim 10^8$\,Ohm
(in the metallic range of densities). 
Only for high densities $n \gtrsim 20\times 10^{11}$cm$^{-2}$, the delay time
becomes comparable to $RC = 10^{-6}$\,s.
The huge time $\tau $ can not be attributed to 
contact phenomena, because for the more disordered sample Si-52, 
at much higher densities (and 
for lower resistance of the contacts, correspondingly), $\tau $ is larger
by a factor of $10^5$. The irrelevance of the contacts
to the time-dependent 
effects is also confirmed  by the 
linearity of the $I-V$-curves measured between 
different contacts with currents in the range 
from $10^{-7}$ down to $10^{-12}$\,A.

{\bf Model.} 
Interface defect charges originating from the lack of stoichiometry
are intrinsic to Si/SiO$_2$ system; their
typical density is 10$^{12}$cm$^{-2}$ for a state-of-art 
thermally grown dioxide \cite{hori}. 
Tunneling of electrons from Si to the interface charged states 
is known to cause a time lag in Si-MOS capacitors at room temperature
\cite{hori}. These charged states are partly 
neutralized during slow cooling of  Si-MOSFETs with a positive gate voltage applied.
Further, at liquid helium temperatures, electron 
tunneling rate to these interface traps in SiO$_2$ 
under a barrier of 3.2\,eV is negligible.  
The uniform part of the potential produced by interface-state charge 
is out of importance,
however, spatial fluctuations of the built-in charge produce
shallow fluctuations of the potential acting on 2D electrons in Si
(at $z>0$), and cause corresponding localization of electrons.
We assume that  the observed times are due
to tunneling processes, on the Si-side entirely, 
between the electrons in 2D ``bulk''  and the potential traps in 
the localized areas produced by the fluctuations of the interface charge
 \cite{fuks}.
For the discussed low-temperature case, the active traps are created 
by the attractive (positive) charges that fall inside a large-scale repulsive 
fluctuation. The attractive charge placed at the interface 
(at $z<0$) localizes an electron nearby
(on the Si-side, $z>0$), with a binding energy \cite{fuks}
\begin{equation}
\varepsilon_b = -m^* e^4/8\kappa^2 \hbar^2.
\end{equation}
Here $m^*=0.21m_e$ is the electron effective mass,
$\kappa = 7.7$ the average dielectric  permittivity,
and  thus $ \varepsilon_b =0.02$\,eV for the Si/SiO$_2$ interface. 
The repulsive charges which are very closely located to 
the attractive charge, decrease the binding electron energy. 
As a result, the binding energy distribution 
broadens and extends over a wide energy range, from about
$\varepsilon_b$ down to 0.
The effective binding energy $\varepsilon_b^{eff}$ 
is therefore substantially lower than $\varepsilon_b$. 

The localized state
is located inside a large-scale  repulsive fluctuation and is
surrounded by a broad potential barrier of the height 
$\varepsilon_b^{eff}$. The barrier itself
is surrounded by the electrons in the metallic regions of the 
2D bulk.  The barrier is responsible for the large
electron capture and emission times. 
At nonzero $T$, only the traps  located close to 
the Fermi level, within $E_F \pm kT$, 
are recharging when the local potential 
in the 2D bulk varies. 

For low temperatures,
the electron emission time \cite{fuks} equals to 
\begin{equation}
\tau_{em}  \approx (\hbar/\varepsilon_b) \exp(-x/\lambda), 
\end{equation}
where $\lambda =\sqrt{\hbar^2/8m^* \varepsilon_b^{eff}}$ is the typical tunneling length
and $x$ is the distance from the trap to the nearest conductive region.
The capture time $\tau_c$ is  related to $\tau_{em} $ 
by the obvious relationship: $\tau_c/\tau_{em} = (1-f)/f$, where 
$f$ is the level occupancy (Fermi distribution function). 
For the traps whose energy is close to $E_F$, $f \approx 1/2$
and these two times are about equal. 
Using, for an estimate, $\varepsilon_b^{eff} = 0.01$eV
we obtain $\lambda = 20$\AA~and the tunneling distance $x= 460$\AA~
corresponding to the tunneling time $10^{-3}$\,s. 
Thus, a radius of the 
repulsive barrier $r $ is to be of the order of 
$500$\AA, to account for the recharging  time of 1\,ms. 
For more disordered samples, the amount of the charge trapped at the interface
and the amplitude of potential fluctuations are even larger.
Therefore, the radius (in the $x-y$ plane) of the potential fluctuations
is also larger. To account for $\tau = 100$s, we estimate $x$ has 
to be equal to 700\AA.
The length scale, (500 - 700)\AA, of the potential fluctuation
is consistent with numerous data obtained for
similar samples, as well as with direct
tunneling microscopy of the Si/SiO$_2$ interface \cite{STM}.

As electron density and $E_F$ increase, 
the potential barriers get thinner \cite{fuks} due to 
screening by free electrons of the 2D bulk, and the radius of the 
total repulsive fluctuation decreases,
causing $\tau $ to decrease monotonically.
This is in a competition with a
density dependence of the effective binding energy $\varepsilon_b^{eff}$;
as a result, the overall density dependence of $\tau $ may 
be non-monotonic.
Finally, above a certain density, 
all localized states sink below the Fermi energy,
the barriers are screened entirely, and
the Drude-Boltzmann regime sets in replacing tunneling. 
The onset occurs at Fermi energy equal roughly to the 
amplitude of bare fluctuations, and is thus inversely proportional 
to the sample peak mobility.

The temperature comes into the model via 
(i) the tunneling distance to the 
nearest trap level found within the energy 
interval $E_F \pm kT$, (ii) Fermi distribution function $f$ and 
(iii) activation processes on the border and inside the localized area.
These mechanisms lead to temperature dependences 
$\tau \propto \exp(-T_{0,i}/T)$, and
the resulting temperature dependence may have different $T_{0,i}$ 
in different ranges of temperature.

The electron tunneling time is much larger than
the transport scattering time ($\approx 3$\,ps), 
the electron-electron interaction time, 
$h /E_{ee} \sim 0.1$\,ps, and the 
electron diffusion time 
($\approx 20$\,ps). Therefore, all electrons in 2D layer
do participate in tunneling during a  
time $\tau \gtrsim 10^{-3}$\,s. 
The capacitance \cite{NDOS} and  Hall voltage 
\cite{JETPL99a} are measured at frequencies lower than  
the tunneling rate, $ 1/\tau_{em} $, 
hence, all the electrons of the 2D bulk participate
in re-charging or in Hall transport.

In summary, our data show that the charge 
transport in the ``metallic conduction'' regime 
is accompanied by (or includes) a non-diffusive component.
It manifests itself in the time-dependent effects
in metallic phase over density range from $n_c$ to about 6 times $n_c$.
The delay/advance time between the ac voltage and the current
in  high mobility samples  is of the ms-range 
and grows exponentially as density and temperature decrease.
For more disordered samples
the time lag between the gate voltage pulse and the response 
reaches 1-100 seconds.
We associate these times with the ``in-plane'' 
tunneling of carriers between the 2D bulk and the potential traps.
The suggested model presumes that the unit ''slow'' trap consists 
of a potential well surrounded by potential barrier,
and  seems to be applicable to various material systems 
since similar long-range localized areas were found in GaAs/AlGaAs as well
\cite{barjoseph}.
Particularly, this may be relevant to the system with 
a set of artificial quantum dots (traps) \cite{ensslin}. 
The model may qualitatively describe (i) large delay time $\tau $, 
(ii) the growth of $\tau$ as temperature and density decrease, and
(iii) the non-monotonic density and temperature dependence of $\tau $.
Whereas the tunneling time in the above model is different from
that in Ref.~\cite{maslov99}, the physics 
of the temperature dependence of the resistivity caused by ''fast traps'' 
(located  nearby the border between the localized and free carriers)
may be similar. The complexity of the presented data, however, requires
a thorough theoretical consideration, which should 
take into account
interaction of the free carriers 
in 2D bulk with the localized ones, 
recharging and  screening processes within 
the localized areas, and symmetry properties of the 
surface localized states.

V.P. acknowledges help by M.\ D'Iorio and E.\ M.\ Goliamina in
samples processing, and stimulating discussions with B.\ Altshuler, D.\
Maslov, and A.\ Finkelstein. 
The work was supported by RFBR, by the Programs ``Physics of
solid-state nanostructures'' and ``Statistical physics'', 
by INTAS, NWO, and
by FWF P13439 and GME, Austria.

\newpage

\end{multicols}

\end{document}